\def\1{\'{\i}}
\def\2{\~n}

\def\b{\beta}
\def\g{\gamma}

\def\ba{\begin{eqnarray}}
\def\ea{\end{eqnarray}}
\def\be{\begin{equation}}
\def\ee{\end{equation}}
\def\beq{\begin{equation}}
\def\eeq{\end{equation}}

\newtheorem{theorem}{Theorem}

\documentclass[preprint,12pt]{elsarticle}

\usepackage{amssymb}
\usepackage{times}

\journal{Physics Letters A}

\begin{document}

\begin{frontmatter}



\title{Symmetries and the compatibility condition for the new translational
shape invariant potentials}


\author{Arturo Ramos}

\address{
Departamento de An\'alisis Econ\'omico,
Universidad de Zaragoza, \\ Gran V\'{\i}a 2, E-50005 Zaragoza, Spain}
\ead{aramos@unizar.es}

\begin{abstract}
In this letter we study a class of symmetries of
the new translational extended shape invariant potentials.
It is proved that a generalization
of a compatibility condition introduced in a previous article
is equivalent to the usual shape invariance condition.
We focus on the recent examples of Odake and Sasaki
(infinitely many polynomial, continuous $l$ and
multi-index rational extensions).
As a byproduct, we obtain new relations,
to the best of our knowledge,
for Laguerre, Jacobi polynomials
and (confluent) hypergeometric functions.
\end{abstract}


\end{frontmatter}



\section{Introduction}

The list of shape invariant potentials has remained quite the same
until 2008. Then, key contributions of G\'omez-Ullate et al.
led to a quick and strong development of the subject in recent years.
The first steps were the possibility of rationally extend
shape-invariant potentials
(to obtain non shape invariant ones) \cite{GomKamMil04,GomKamMil04b}.
Then, the introduction of the so called $X_{l}$
exceptional Laguerre and Jacobi polynomials
\cite{GomKamMil10,GomKamMil09} fostered all subsequent works.
By the one hand, Quesne (and coworkers) \cite{Que08,Que09,BagQueRoy09}
introduced the first examples of rationally extended shape invariant potentials. This idea has been greatly developed by
Odake and Sasaki \cite{OdaSas09,OdaSas10,OdaSas10b,OdaSas11,OdaSas11b}
to infinitely many families of rationally extended shape-invariant potentials, even with functions depending on continuous
index ${l}$ and multi-indexed polynomials. They have also
extended these
ideas to the context of discrete quantum mechanics
(see, e.g., \cite{OdaSas10c} and references therein).
Other works by Grandati \cite{Gra11,Gra12,Gra12b}
have a close relation with the ones cited.

On the other hand, the works
\cite{Que08,BouGanMal10,BouGanMal11} inspired our recent article
\cite{Ram11}, where a compatibility condition has been found
that is satisfied by the new examples. Even we have shown
that such a condition forces the shape invariance of the examples
treated there. It is worth mentioning that \cite{BouGanMal10,BouGanMal11}
are preceded by \cite{GanMal08}.
This paper continues on the same line of study and shows that
the examples of \cite{OdaSas09,OdaSas10,OdaSas10b,OdaSas11}
fit perfectly in our framework, satisfying the mentioned
compatibility condition.

The letter is organized as follows. In the second
section we recall the equations which satisfy the new
translational shape invariant potentials of
\cite{Que08,Que09,BagQueRoy09,BouGanMal11,Ram11}.
We prove the equivalence between a generalization of the
cited compatibility condition and the usual shape invariance condition.
Afterwards, we comment on the isospectrality properties
of the potentials involved.
In the third section we describe how the examples of
\cite{OdaSas09,OdaSas10,OdaSas10b,OdaSas11} fit
into our framework. We obtain as a byproduct new relations,
to the best of our knowledge,
for Laguerre, Jacobi polynomials and (confluent)
hypergeometric functions.
In the fourth and last section we offer some conclusions.

\section{Symmetries and the relation of the compatibility
condition with the shape invariance condition\label{eccsic}}

For a brief account of shape invariance, 
see, e.g., \cite{Ram11} and references therein.
In the examples
of \cite{Que08,Que09,BagQueRoy09,OdaSas09,OdaSas10,OdaSas10b,OdaSas11,BouGanMal11,Ram11}
the superpotential function takes the form of
\be
W(x,a)=W_0(x,a)+W_{1+}(x,a)-W_{1-}(x,a)\,, \label{Wgen}
\ee
where $a$ denotes the set of parameters under transformation.
$W_0(x,a)$ is the superpotential of a pair of
shape invariant partner potentials of the classical type.
$W_{1+}(x,a)$, $W_{1-}(x,a)$ are logarithmic
derivatives which moreover satisfy
\begin{equation}
W_{1-}(x,a)=W_{1+}(x,f(a))\,, \label{sic2}
\end{equation}
where $f(a)$ in those cases is a translation of $a$.

The corresponding partner potentials for (\ref{Wgen})
are
\ba
V(x,a)&=&W_0^2(x,a)-W_0^{\prime}(x,a)   \nonumber\\
& &+W_{1+}^2(x,a)+W_{1+}^{\prime}(x,a)
+W_{1-}^2(x,a)+W_{1-}^{\prime}(x,a)     \nonumber\\
& &-2W_0(x,a)W_{1-}(x,a)+2W_0(x,a)W_{1+}(x,a) \nonumber\\
& &-2W_{1-}(x,a)W_{1+}(x,a)-2W_{1+}^\prime(x,a)   \label{Vcom}\\
\widetilde V(x,a)&=&W_0^2(x,a)+W_0^{\prime}(x,a)   \nonumber\\
& &+W_{1+}^2(x,a)+W_{1+}^{\prime}(x,a)
+W_{1-}^2(x,a)+W_{1-}^{\prime}(x,a)     \nonumber\\
& &-2W_0(x,a)W_{1-}(x,a)+2W_0(x,a)W_{1+}(x,a) \nonumber\\
& &-2W_{1-}(x,a)W_{1+}(x,a)-2W_{1-}^\prime(x,a)   \label{Vtilcom}
\ea
However, for the examples
of \cite{Que08,Que09,BagQueRoy09,BouGanMal11} such partner
potentials reduce to
\ba
V(x,a)&=&V_0(x,a)-2W_{1+}^{\prime}(x,a)\,,                  \label{Vg}        \\
\widetilde V(x,a)&=&\widetilde V_0(x,a)-2W_{1-}^\prime(x,a)\,,    \label{tilVg}
\ea
where $V_0(x,a)$, $\widetilde V_0(x,a)$ conform
the pair of shape invariant partner potentials
associated to $W_0(x,a)$. Thus, it is in principle
necessary that the following \emph{compatibility condition}
holds:
\begin{equation}
W_{1+}^2+W_{1+}^{\prime}+W_{1-}^2+W_{1-}^{\prime}
-2W_0W_{1-}+2W_0W_{1+}-2W_{1-}W_{1+}=0 \label{cc1}
\end{equation}
(the dependence on the arguments has been omitted for brevity).
Such compatibility condition is the main object of our interest here.
First we will discuss a kind of symmetries of
the problems of type (\ref{Wgen}), (\ref{sic2}), (\ref{Vcom}), (\ref{Vtilcom}).
Afterwards we establish the relation between a generalized compatibility
condition and the ordinary shape invariance condition.

\subsection{Symmetries of the new translational shape invariant potentials\label{symt}}

There exist a class of symmetries of superpotentials of
type (\ref{Wgen}) which satisfy the condition (\ref{sic2})
given by the transformations
\ba
W_{1+}(x,a)&=&U_{1+}(x,a)-g(x) \label{trW1p}\\
W_{1-}(x,a)&=&U_{1-}(x,a)-g(x) \label{trW1m}
\ea
where $g(x)$ is a function \emph{depending only on $x$}.
The function $g(x)$ must be differentiable in the domain 
of interest but otherwise arbitrary.
For example, $g(x)$ could be any polynomial, ${\rm e}^x$, etc.
Thus we have
$$
W(x,a)=W_0(x,a)+W_{1+}(x,a)-W_{1-}(x,a)
=W_0(x,a)+U_{1+}(x,a)-U_{1-}(x,a)
$$
The corresponding partner
potentials (\ref{Vcom}), (\ref{Vtilcom}) are
likewise invariant under (\ref{trW1p}) and (\ref{trW1m}).
However, their different terms do vary, in such a way
that their variations cancel out. Firstly, we have
\ba
& &W_{1+}^2+W_{1+}^{\prime}+W_{1-}^2+W_{1-}^{\prime}
-2W_0W_{1-}+2W_0W_{1+}-2W_{1-}W_{1+}    \nonumber\\
& &=U_{1+}^2+U_{1+}^{\prime}+U_{1-}^2+U_{1-}^{\prime}
-2W_0U_{1-}+2W_0U_{1+}-2U_{1-}U_{1+}     \nonumber\\
& &\quad-2g^\prime(x)                     \nonumber   
\ea
and moreover
\ba
-2W_{1+}^\prime(x,a)&=&-2U_{1+}^\prime(x,a)+2g^\prime(x) \nonumber\\
-2W_{1-}^\prime(x,a)&=&-2U_{1-}^\prime(x,a)+2g^\prime(x) \nonumber
\ea
Therefore, if (\ref{cc1}) holds, we have
\ba
& &U_{1+}^2(x,a)+U_{1+}^{\prime}(x,a)
+U_{1-}^2(x,a)+U_{1-}^{\prime}(x,a) \nonumber\\
& &-2W_0(x,a)U_{1-}(x,a)+2W_0(x,a)U_{1+}(x,a)-2U_{1-}(x,a)U_{1+}(x,a)
=2g^\prime(x)                       \nonumber
\ea
This means that by virtue of a symmetry of the problem, the compatibility
condition (\ref{cc1}) should be generalized in such a way that
its right hand side could be a function of $x$ not necessarily equal to zero.
This observation leads to our main result in the following subsection.

\subsection{Compatibility and shape invariance conditions}

For the class of problems described in this letter, there is
an equivalence between the mentioned generalized compatibility
condition and the usual shape invariance condition,
as described in the next Theorem.

\begin{theorem}
Assume we have a superpotential of the type
\be
W(x,a)=W_0(x,a)+W_{1+}(x,a)-W_{1-}(x,a)\,,
\ee
where
$$
W_{1-}(x,a)=W_{1+}(x,f(a))\,,
$$
$f(a)$ being the transformation on the parameters $a$,
and $W_0(x,a)$ satisfies the shape invariance condition
\begin{equation}
W_0^2(x,a)-W_0^2(x,f(a))+W_0^\prime(x,f(a))+W_0^\prime(x,a)=R(f(a)). \label{siW0}
\end{equation}
Then, the shape invariant condition for $W(x,a)$
\be
W^2(x,a)-W^2(x,f(a))+W^\prime(x,f(a))+W^\prime(x,a)=R(f(a))\label{siWt}
\ee
holds if and only if
\ba
& &W_{1+}^2(x,a)+W_{1+}^{\prime}(x,a)+W_{1-}^2(x,a)+W_{1-}^{\prime}(x,a)\nonumber\\
& &-2W_0(x,a)W_{1-}(x,a)+2W_0(x,a)W_{1+}(x,a)-2W_{1-}(x,a)W_{1+}(x,a)\nonumber\\
& &=\epsilon(x) \label{cidelta}
\ea
for some non-singular function $\epsilon(x)$ of $x$ only.
\end{theorem}

{\it Proof}

The condition of shape invariance (\ref{siWt}) reads in this case
\begin{eqnarray}
& &W^2(x,a)-W^2(x,f(a))+W'(x,f(a))+W'(x,a)-R(f(a))=\nonumber\\
& &\quad W_0^2(x,a)-W_0^2(x,f(a))+W_0'(x,f(a))+W_0'(x,a)-R(f(a))\nonumber\\
& &\quad+W_{1+}^2(x,a)+W_{1+}^{\prime}(x,a)
+W_{1-}^2(x,a)+W_{1-}^{\prime}(x,a) \nonumber\\
& &\quad-2W_0(x,a)W_{1-}(x,a)+2W_0(x,a)W_{1+}(x,a)
-2W_{1-}(x,a)W_{1+}(x,a)\nonumber\\
& &\quad -[W_{1+}^2(x,f(a))+W_{1+}^{\prime}(x,f(a))
+W_{1-}^2(x,f(a))+W_{1-}^{\prime}(x,f(a)) \nonumber\\
& &\quad -2W_0(x,f(a))W_{1-}(x,f(a))+2W_0(x,f(a))W_{1+}(x,f(a))\nonumber\\
& &\quad -2W_{1-}(x,f(a))W_{1+}(x,f(a))]-2W_{1-}^{\prime}(x,a)
+2W_{1+}^{\prime}(x,f(a))=0\nonumber\\
& & \label{gor}
\end{eqnarray}
With the hypothesis that $W_0(x,a)$ satisfies (\ref{siW0}),
also that $W_{1-}(x,a)=W_{1+}(x,f(a))$ and that
\ba
& &W_{1+}^2(x,a)+W_{1+}^{\prime}(x,a)+W_{1-}^2(x,a)+W_{1-}^{\prime}(x,a)\nonumber\\
& &-2W_0(x,a)W_{1-}(x,a)+2W_0(x,a)W_{1+}(x,a)-2W_{1-}(x,a)W_{1+}(x,a)\nonumber\\
& &=\epsilon(x)\nonumber\\
& &W_{1+}^2(x,f(a))+W_{1+}^{\prime}(x,f(a))+W_{1-}^2(x,f(a))
+W_{1-}^{\prime}(x,f(a))\nonumber\\
& &-2W_0(x,f(a))W_{1-}(x,f(a))+2W_0(x,f(a))W_{1+}(x,f(a))\nonumber\\
& &-2W_{1-}(x,f(a))W_{1+}(x,f(a))\nonumber\\
& &=\epsilon(x)\nonumber
\ea
the shape invariance condition is readily satisfied.

Conversely, with the above hypothesis we assume that the
shape invariance condition (\ref{siWt}) is satisfied,
therefore (\ref{gor}) is also satisfied.
Taking into account (\ref{siW0})
and $W_{1-}(x,a)=W_{1+}(x,f(a))$ and rearranging,
(\ref{gor}) becomes
\begin{eqnarray}
& &\quad W_{1+}^2(x,a)+W_{1+}^{\prime}(x,a)
+W_{1-}^2(x,a)+W_{1-}^{\prime}(x,a) \nonumber\\
& &\quad-2W_0(x,a)W_{1-}(x,a)+2W_0(x,a)W_{1+}(x,a)\nonumber\\
& &\quad -2W_{1-}(x,a)W_{1+}(x,a)=\nonumber\\
& &\quad W_{1+}^2(x,f(a))+W_{1+}^{\prime}(x,f(a))
+W_{1-}^2(x,f(a))+W_{1-}^{\prime}(x,f(a)) \nonumber\\
& &\quad -2W_0(x,f(a))W_{1-}(x,f(a))+2W_0(x,f(a))W_{1+}(x,f(a))\nonumber\\
& &\quad -2W_{1-}(x,f(a))W_{1+}(x,f(a))\nonumber
\end{eqnarray}
that is, the expression evaluated at $(x,a)$ equals the
expression itself evaluated at $(x,f(a))$, thus both expressions
must be equal to a function of $x$ only, namely, $\epsilon(x)$.
This ends the proof of the Theorem.

{\it Remarks\/}

1. In actual examples it is observed that
(\ref{cidelta}) is satisfied with $\epsilon(x)=0$, which
is a slightly stronger condition that in particular
implies shape invariance.

2. Note that Ho proposes in \cite{Ho11,Ho11b} a similar
form to the superpotential (\ref{Wgen}), but
that approach is different: other relations, different
from (\ref{cc1}) or (\ref{cidelta}) are satisfied.
As an example of this,
in \cite{Ram11} it is shown that the harmonic oscillator
and the Morse potential admit no non-trivial extensions
by our means. However with the technique of Ho they do.
See also \cite{CarPerRanSan08,FelSmi09,Gra11b,Que12}.

3. We observe that the potentials in (\ref{Vg}) and (\ref{tilVg})
are related by construction by a first order intertwining
relation as described in Section~2 of \cite{Ram11} with superpotential
(\ref{Wgen}). The fulfillment of condition (\ref{cc1})
provides a cancelation of some of their terms (it is the same condition
for $V(x,a)$ in (\ref{Vg}) and $\widetilde V(x,a)$ in (\ref{tilVg})).
Thus the isospectrality (maybe up to the ground
state of one of them) of the potentials
(\ref{Vg}) and (\ref{tilVg}) is ensured.
See \cite{CarFerRam01,CarRam08} for a group theoretical
explanation of the intertwining technique.

4. Another question is the
isospectrality of the mentioned potentials with the ordinary
shape invariant potentials $V_0(x,a)$ and $\widetilde V_0(x,a)$.
This is also easy to justify: with the conditions (\ref{sic2})
and (\ref{cidelta}) the shape invariant relation (\ref{siWt})
for the potentials of (\ref{Vg}) and (\ref{tilVg}) becomes
identical to that of the partner potentials $V_0(x,a)$
and $\widetilde V_0(x,a)$. In particular, the quantity $R(f(a))$,
{}from which the spectrum of the potentials is calculated, is
identical in both cases, showing the mentioned isospectrality
(maybe up to the ground state of one of them). See also
\cite{BagQueRoy09,OdaSas11} for an approximation to
such an isospectrality based on the intertwining technique.

5. We have established that
the generalized compatibility condition (\ref{cidelta}) is equivalent to
the ordinary shape invariance condition (\ref{siWt}) in the mentioned
circumstances.
However, the former condition is simpler to
work with than the latter for the cases studied in
\cite{Que08,Que09,BagQueRoy09,BouGanMal11,Ram11}
and in this letter.

6. The condition (\ref{cc1}) has been shown, for $W's$
satisfying the Bernoulli equation $W^\prime+W^2-k_1(x)W=0$
(where $k_1(x)=c\coth (c x)$, etc.),
in the examples of \cite{Ram11}, to imply (\ref{sic2}) in particular.
This means that such conditions are not really
independent in specific examples.

The compatibility condition (\ref{cidelta}) admits another, even simpler, form. Denoting
$$
W_{1+}(x,a)=\frac{\psi^\prime_{1+}(x,a)}{\psi_{1+}(x,a)}\,,\quad\quad\quad
W_{1-}(x,a)=\frac{\psi^\prime_{1-}(x,a)}{\psi_{1-}(x,a)}
$$
it becomes
\begin{equation}
\frac{1}{\psi_{1+}}(\psi^{\prime\prime}_{1+}+2W_0 \psi^{\prime}_{1+})
+\frac{1}{\psi_{1-}}(\psi^{\prime\prime}_{1-}-2W_0\psi^{\prime}_{1-})
-2\frac{\psi^{\prime}_{1+}\psi^{\prime}_{1-}}{\psi_{1+}\psi_{1-}}=\epsilon(x)
\label{cc12d}
\end{equation}
where the dependence on the
arguments $(x,a)$ has been dropped for simplicity.
For the case of $\epsilon(x)=0$ it follows
\begin{equation}
(\psi^{\prime\prime}_{1+}+2W_0 \psi^{\prime}_{1+})\psi_{1-}
+(\psi^{\prime\prime}_{1-}-2W_0\psi^{\prime}_{1-})\psi_{1+}
-2\psi^{\prime}_{1+}\psi^{\prime}_{1-}=0
\label{cc12}
\end{equation}

In terms of the functions $\psi_{1+}(x,a), \psi_{1-}(x,a)$, the
symmetries of Subsection~\ref{symt} are expressed in the following
way. The functions change as
\ba
\psi_{1+}(x,a)&=&\exp\left(-\int g(x)\,dx\right)\chi_{1+}(x,a) \nonumber\\
\psi_{1-}(x,a)&=&\exp\left(-\int g(x)\,dx\right) \chi_{1-}(x,a) \nonumber
\ea
where
${\displaystyle U_ {1+}(x,a)=\frac{\chi_{1+}^\prime(x,a)}{\chi_{1+}(x,a)}}$
and ${\displaystyle U_ {1-}(x,a)=\frac{\chi_{1-}^\prime(x,a)}{\chi_{1-}(x,a)}}$.

\section{Examples \label{examples}}

In this section we study the fulfillment
of the compatibility condition (\ref{cc12})
for the examples of \cite{OdaSas09,OdaSas10,OdaSas10b,OdaSas11}.
These cases are specially well suited for our purposes,
since they take the form of Section~\ref{eccsic} and are
known to be shape invariant. By the symmetry property of these problems,
it suffices to study the compatibility condition (\ref{cc12}),
which will be obtained directly in all cases.
We will obtain as a byproduct new relations, to the best of our knowledge,
of Laguerre, Jacobi polynomials and (confluent) hypergeometric functions.

\subsection{Polynomial shape invariant extensions
of the radial oscillator and Darboux--P\"oschl--Teller potentials}

\subsubsection{Radial oscillator}

According to \cite{OdaSas09,OdaSas10,OdaSas10b}, the extended partner
potentials of the radial oscillator have a superpotential of the form
$$
W_l(x,g)=W_0(x,g+l)+\frac{\xi^\prime_l(x^2,g+1)}{\xi_l(x^2,g+1)}
-\frac{\xi^\prime_l(x^2,g)}{\xi_l(x^2,g)}
$$
where $x>0$,
\begin{eqnarray}
W_0(x,g)&=&-x+\frac{g}{x}       \nonumber\\
\xi_l(x,g)&=&L_l^{(g+l-\frac 32)}(-x)\nonumber
\end{eqnarray}
and $L_n^{(a)}(x)$ are Laguerre polynomials.

We will try to check (\ref{cc12}) directly choosing
(with a slight abuse of notation)
\begin{eqnarray}
W_0(x,a)&=&W_0(x,g+l) \nonumber\\
\psi_{1+}(x,a)&=&\xi_l(x^2,g+1)\nonumber\\
\psi_{1-}(x,a)&=&\xi_l(x^2,g)\nonumber
\end{eqnarray}
and by writing it in another way, using
the relation (2.41) of \cite{OdaSas10b}, namely
(dependence on arguments dropped)
\begin{eqnarray}
\psi^{\prime\prime}_{1+}&=&
4l\psi_{1+}-2\left(\frac{g+l}{x}+x\right)\psi^{\prime}_{1+}\label{pp_RO}\\
\psi^{\prime\prime}_{1-}&=&
4l\psi_{1-}-2\left(\frac{g+l-1}{x}+x\right)\psi^{\prime}_{1-}\label{pm_RO}
\end{eqnarray}
Thus the relation (\ref{cc12}) becomes
\begin{equation}
8l \psi_{1+}\psi_{1-}+\frac{2(1-2g-2l)}{x}\psi_{1+}\psi^\prime_{1-}
-4x\psi_{1-}\psi^\prime_{1+}-2\psi^\prime_{1-}\psi^\prime_{1+}=0 \label{cc12RO}
\end{equation}
This last relation can be proved using the equations (3.5) and (3.6) of \cite{OdaSas10b}. That implies, in particular, the fulfillment of
(\ref{siWt}) for this case. In \cite{OdaSas10b} it is
proved (\ref{siWt}) directly for the current case.

The relations (\ref{cc12}) and (\ref{cc12RO}) are new,
and equivalent to each other, for Laguerre polynomials.

\subsubsection{Trigonometric Darboux-P\"oschl-Teller potential}

According to \cite{OdaSas09,OdaSas10,OdaSas10b}, the extended
partner potentials of the trigonometric
Darboux-P\"oschl-Teller potential have a superpotential of the form
$$
W_l(x,g,h)=W_0(x,g+l,h+l)
+\frac{\xi^\prime_l(\cos(2x),g+1,h+1)}{\xi_l(\cos(2x),g+1,h+1)}
-\frac{\xi^\prime_l(\cos(2x),g,h)}{\xi_l(\cos(2x),g,h)}
$$
where $x\in\left(0,\frac{\pi}{2}\right)$,
\begin{eqnarray}
W_0(x,g,h)&=&g \cot(x)-h \tan(x)       \nonumber\\
\xi_l(x,g,h)&=&P_l^{(-g-l-\frac 12,h+l-\frac 32)}(x)\nonumber
\end{eqnarray}
and $P_n^{(a,b)}(x)$ are Jacobi polynomials.

We will try to check (\ref{cc12}) by choosing
(with a slight abuse of the notation)
\begin{eqnarray}
W_0(x,a)&=&W_0(x,g+l,h+l) \nonumber\\
\psi_{1+}(x,a)&=&\xi_l(\cos(2x),g+1,h+1)\nonumber\\
\psi_{1-}(x,a)&=&\xi_l(\cos(2x),g,h)\nonumber
\end{eqnarray}
Moreover, we transform (\ref{cc12}) by using
the relation (2.41) of \cite{OdaSas10b}, namely
\begin{eqnarray}
\psi^{\prime\prime}_{1+}&=&
4l(g-h-l+1)\psi_{1+}
+2\left((g+l+1)\cot x+(h+l)\tan x\right)\psi^{\prime}_{1+}\nonumber\\
\psi^{\prime\prime}_{1-}&=&
4l(g-h-l+1)\psi_{1-}
+2\left((g+l)\cot x+(h+l-1)\tan x\right)\psi^{\prime}_{1-}\nonumber
\end{eqnarray}
Thus, the relation (\ref{cc12}) becomes
\begin{eqnarray}
& &-8l(h-g+l-1)\psi_{1+}\psi_{1-}
+2(2h+2l-1)\tan x\,\psi_{1+}\psi^\prime_{1-} \nonumber\\
& &\quad+2(2g+2l+1)\cot x\,\psi_{1-}\psi^\prime_{1+}
-2\psi^\prime_{1-}\psi^\prime_{1+}=0 \label{cc12TPT}
\end{eqnarray}
Such relation can be proved directly using (3.12) and (3.13)
of \cite{OdaSas10b}. In such a paper, it has been proved the shape invariance condition (\ref{siWt}) directly. We have proved it checking that the
stronger (and simpler) condition (\ref{cc12}) or (\ref{cc12TPT}) holds.

For this case, (\ref{cc12}) and (\ref{cc12TPT}) are new
relations, equivalent to each other, for Jacobi polynomials.

\subsubsection{Hyperbolic Darboux-P\"oschl-Teller potential}

According to \cite{OdaSas09,OdaSas10,OdaSas10b}, the extended partner
potentials of the hyperbolic Darboux-P\"oschl-Teller potential have a
superpotential of the form
$$
W_l(x,g,h)=W_0(x,g+l,h-l)
+\frac{\xi^\prime_l(\cosh(2x),g+1,h-1)}{\xi_l(\cosh(2x),g+1,h-1)}
-\frac{\xi^\prime_l(\cosh(2x),g,h)}{\xi_l(\cosh(2x),g,h)}
$$
where $x>0$,
\begin{eqnarray}
W_0(x,g,h)&=&g \coth(x)-h \tanh(x)       \nonumber\\
\xi_l(x,g,h)&=&P_l^{(-g-l-\frac 12,-h+l-\frac 32)}(x)\nonumber
\end{eqnarray}
and $P_n^{(a,b)}(x)$ are again Jacobi polynomials.

We will try to check (\ref{cc12}) by choosing
\begin{eqnarray}
W_0(x,a)&=&W_0(x,g+l,h-l) \nonumber\\
\psi_{1+}(x,a)&=&\xi_l(\cosh(2x),g+1,h-1)\nonumber\\
\psi_{1-}(x,a)&=&\xi_l(\cosh(2x),g,h)\nonumber
\end{eqnarray}
and transforming the cited condition by using
the relation (2.41) of \cite{OdaSas10b}, namely
\begin{eqnarray}
\psi^{\prime\prime}_{1+}&=&
4l(l-g-h-1)\psi_{1+}
+2\left((g+l+1)\coth x+(h-l)\tanh x\right)\psi^{\prime}_{1+}\nonumber\\
\psi^{\prime\prime}_{1-}&=&
4l(l-g-h-1)\psi_{1-}
+2\left((g+l)\coth x+(h-l+1)\tanh x\right)\psi^{\prime}_{1-}\nonumber
\end{eqnarray}
Thus the relation (\ref{cc12}) becomes
\begin{eqnarray}
& &-8l(h+g-l+1)\psi_{1+}\psi_{1-}
+2(1+2h-2l)\tanh x\,\psi_{1+}\psi^\prime_{1-} \nonumber\\
& &\quad+2(1+2g+2l)\coth x\,\psi_{1-}\psi^\prime_{1+}
-2\psi^\prime_{1-}\psi^\prime_{1+}=0 \label{cc12HPT}
\end{eqnarray}
This last relation can be proved by using equations (3.12) and (3.13) of \cite{OdaSas10b}, as in the previous case.
Therefore, the shape invariance for this case holds in particular.
In \cite{OdaSas10b}, the relation (\ref{siWt}) has been proved directly.

For this case, (\ref{cc12}) and (\ref{cc12HPT})
are new relations, equivalent to each other,
for Jacobi polynomials.

\subsection{Continuous $l$ shape invariant extensions of
the radial oscillator and
trigonometric Darboux--P\"oschl--Teller potentials}

\subsubsection{Radial oscillator}

According to \cite{OdaSas11}, the extended partner
potentials of the radial oscillator with continuous $l>0$
have a superpotential of the form
$$
W_l(x,g)=W_0(x,g+l)+\frac{\xi^\prime_l(x^2,g+1)}{\xi_l(x^2,g+1)}
-\frac{\xi^\prime_l(x^2,g)}{\xi_l(x^2,g)}
$$
where $x>0$,
\begin{eqnarray}
W_0(x,g)&=&-x+\frac{g}{x}       \nonumber\\
\xi_l(x,g)&=&\frac{\Gamma(g+2 l-\frac12)}{\Gamma(l+1)\Gamma(g+l-\frac12)}\,
  {}_1F_1\left(\begin{array}{c}-l\\g+l-\frac12\end{array}
  \Bigm|-x\right)\nonumber
\end{eqnarray}
and ${}_1F_1\left(\begin{array}{c}a\\b\end{array}\Bigm|x\right)$,
$\Gamma(x)$ are the confluent hypergeometric and Gamma functions,
respectively.

We choose (with a slight abuse of notation)
\begin{eqnarray}
W_0(x,a)&=&W_0(x,g+l) \nonumber\\
\psi_{1+}(x,a)&=&\xi_l(x^2,g+1)\nonumber\\
\psi_{1-}(x,a)&=&\xi_l(x^2,g)\nonumber
\end{eqnarray}
in order to check whether (\ref{cc12}) is satisfied.
We first transform it by using the relation (3.9)
of \cite{OdaSas11}, namely (\ref{pp_RO}) and (\ref{pm_RO}).
Therefore, (\ref{cc12}) is transformed into (\ref{cc12RO}) again.
Such relation can be proved again for the current $\psi_{1+}, \psi_{1-}$ by
using properties (3.10) and (3.11) of \cite{OdaSas11}.
Thus the compatibility condition holds and as a result also
the shape invariance condition does. This last
result has been obtained
directly in \cite{OdaSas11}.

For this case, (\ref{cc12}) and (\ref{cc12RO}) are new relations,
equivalent to each other, for confluent hypergeometric functions.

\subsubsection{Trigonometric Darboux-P\"oschl-Teller potential}

According to \cite{OdaSas11}, the extended partner
potentials of the trigonometric Darboux-P\"oschl-Teller
potential with continuous $l>0$
have a superpotential of the form
$$
W_l(x,g,h)=W_0(x,g+l,h+l)
+\frac{\xi^\prime_l(\cos(2x),g+1,h+1)}{\xi_l(\cos(2x),g+1,h+1)}
-\frac{\xi^\prime_l(\cos(2x),g,h)}{\xi_l(\cos(2x),g,h)}
$$
where $x\in\left(0,\frac{\pi}{2}\right)$,
\begin{eqnarray}
W_0(x,g,h)&=&g \cot(x)-h \tan(x)       \nonumber\\
\xi_l(x,g,h)&=&
\frac{\Gamma(g+2 l-\frac12)}{\Gamma(l+1)\Gamma(g+l-\frac12)}\,
  {}_2F_1\left(\begin{array}{c}-l, g-h+l-1\\g+l-\frac12\end{array}
  \Bigm|\frac{1-x}{2}\right)\nonumber
\end{eqnarray}
and ${}_2F_1\left(\begin{array}{c}a,b\\c\end{array}\Bigm|x\right)$
is the hypergeometric function.

We denote again
\begin{eqnarray}
W_0(x,a)&=&W_0(x,g+l,h+l) \nonumber\\
\psi_{1+}(x,a)&=&\xi_l(\cos(2x),g+1,h+1)\nonumber\\
\psi_{1-}(x,a)&=&\xi_l(\cos(2x),g,h)\nonumber
\end{eqnarray}
in order to check that (\ref{cc12}) holds.
We transform it first using the result (3.9) of \cite{OdaSas11},
namely
\begin{eqnarray}
\psi^{\prime\prime}_{1+}&=&
-4l(g-h+l-1)\psi_{1+}\nonumber\\
& &-2\left((g+h+2l+1)\csc (2x)+(g-h-1)\cot(2x)\right)\psi^{\prime}_{1+}\nonumber\\
\psi^{\prime\prime}_{1-}&=&
-4l(g-h+l-1)\psi_{1-}\nonumber\\
& &-2\left((g+h+2l-1)\csc (2x)+(g-h-1)\cot(2x)\right)\psi^{\prime}_{1-}\nonumber
\end{eqnarray}
Thus the relation (\ref{cc12}) becomes
\begin{eqnarray}
& & -8l(g-h+l-1)\psi_{1+}\psi_{1-}
-2(2g+2l-1)\cot x\,\psi_{1+}\psi^\prime_{1-} \nonumber\\
& &\quad-2(2h+2l+1)\tan x\,\psi_{1-}\psi^\prime_{1+}
-2\psi^\prime_{1-}\psi^\prime_{1+}=0 \label{cc12T2F1PT}
\end{eqnarray}
This last equation can be proved directly using the
equations (3.10) and (3.11) of \cite{OdaSas11},
thus fulfilling the compatibility condition.
As a consequence, (\ref{siWt}) holds
(something which has been
checked directly in \cite{OdaSas11}).

For this case, (\ref{cc12}) and (\ref{cc12T2F1PT}) are
new relations amongst hypergeometric functions
equivalent to each other.

\section{Conclusions and outlook}

We have studied the fulfillment of the compatibility
condition introduced in \cite{Ram11} in the cases of
the extended shape invariant potentials of
\cite{OdaSas09,OdaSas10,OdaSas10b,OdaSas11}.
Firstly, we have proved that for the form of the
superpotential (\ref{Wgen}),
where $W_0(x,a)$
generates a pair of shape invariant potentials of
the classical type and the extra terms satisfy
(\ref{sic2}), the compatibility condition
(\ref{cidelta}) is equivalent to the ordinary
shape invariance condition for the full
superpotential (\ref{siWt}).
Then, the cited examples are exactly
of the form described in Section~\ref{eccsic}.
We check directly whether the compatibility condition
(\ref{cc12}) holds and indeed we prove it in all cases,
using previous results of \cite{OdaSas10b,OdaSas11}.
Thus, for the cases studied we provide an
alternative and simpler way of proving shape invariance.

The multi-index polynomial extensions to the radial oscillator and
trigonometric Darboux-P\"oschl-Teller potentials introduced in
\cite{OdaSas11b} are shown to be shape
invariant and they are of the form described in
Section~\ref{eccsic}, thus the compatibility
condition (\ref{cidelta}) must hold in that cases as well.

It would be interesting to see whether there exists non-trivial rational extensions to other shape invariant potentials of the Infeld and Hull classification \cite{InfHul51,CarRam00,CooKhaSuk01,GanMalRas11} (with superpotential of the type $k_0(x)+m k_1(x)$) to infinitely many polynomial and continuous $l$ functions analogous
to that of \cite{OdaSas09,OdaSas10,OdaSas10b,OdaSas11}. If these examples do exist, the relation (\ref{cidelta}) must hold again.

\section*{Acknowledgements}
We acknowledge correspondence with R. Sasaki, where
he advanced us the fulfillment of relation (\ref{cc1})
for some of their cases and made helpful remarks in
a previous version of this paper. This work is
supported by Spanish Ministry of Economy and Competitiveness,
project ECO2009-09332 and by Aragon Government,
ADETRE Consolidated Group.













\end{document}